\documentclass{article}

\usepackage{arxiv}

\usepackage[utf8]{inputenc} 
\usepackage[T1]{fontenc}    
\usepackage{hyperref}       
\usepackage{url}            
\usepackage{booktabs}       
\usepackage{amsfonts}       
\usepackage{nicefrac}       
\usepackage{microtype}      
\usepackage{lipsum}		
\usepackage{graphicx}
\usepackage{natbib}
\usepackage{doi}

\usepackage{ltablex}

\title{Intelligence and Global Bias in the Stock Market}

\author{ \href{https://orcid.org/0000-0001-7193-6736}{\includegraphics[scale=0.06]{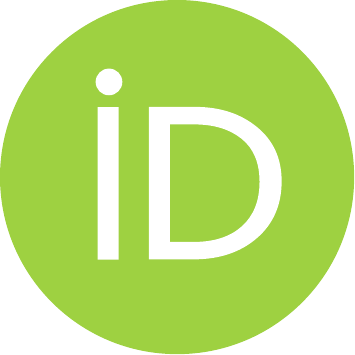}\hspace{1mm}Kazuo Sano}\thanks{	\texttt{sano@fpu.ac.jp}} \\
\\
	Department of Economics\\
	Fukui Prefectural University\\
	Fukui 910-1195, Japan \\
 \\
	SOBIN Institute LLC\\
	3-38-7 Keyakizaka, Kawanishi\\
	Hyogo 666-0145, Japan
}

\renewcommand{\shorttitle}{\textit{arXiv} Template}

\hypersetup{
pdftitle={Intelligence and Global Bias in the Stock Market},
pdfsubject={q-bio.NC, q-bio.QM},
pdfauthor={Kazuo Sano},
pdfkeywords={prospect theory; value function; cusp; anterior cingulate cortex (ACC); lateral habenula (LHb); orbito-frontal cortex (OFC); dorsal striatum (DS); ventral striatum (VS)},
}

\begin{document}

\maketitle

\begin{abstract}
Trade is one of the essential feature of human intelligence. The securities market is the ultimate expression of it. The fundamental indicators of stocks include information as well as the effects of noise and bias on the stock prices; however, identifying the effects of noise and bias is generally difficult. In this article, I present the true fundamentals hypothesis based on rational expectations and detect the global bias components from the actual fundamental indicators by using a log-normal distribution model based on the true fundamentals hypothesis. The analysis results show that biases generally exhibit the same characteristics, strongly supporting the true fundamentals hypothesis. Notably, the positive price-to-cash flows from the investing activities ratio is a proxy for the true fundamentals. Where do these biases come from? The answer is extremely simple: ``Cash is a fact, profit is an opinion.'' Namely, opinions of management and accounting are added to true fundamentals. As a result, Kesten process is realized and the Pareto distribution is to be obtained. This means that the market knows it and represents as a stable global bias in the stock market.
\end{abstract}


\section{Introduction}

Can cats or dogs do barter? Trade is one of the essential feature of human intelligence. \cite{10.1371/journal.pone.0001518} reported that chimpanzees who are an intelligent species like human have the ability to trade, but are reluctant to trade. This subsequently led to a major divergence in the fates of the two species. The market is nothing but an expression of human intelligence. Intelligence does not arise only in individual brains; it also arises in groups of individuals~(\cite{malone2015handbook}). The securities market, including crypto-assets, is the ultimate expression of human intelligence. As considered by \cite{fama1965behavior}, if the security market is efficient in the strong or semi-strong sense, information on securities instantaneously changes the traders’ subjective equilibrium and the differences in the speed with which they respond to the information decides who are the winners and losers. If the security market is efficient in the weak sense, the market equilibrium trends according to changes in the traders’ subjective equilibrium because of noise. The efficiency of security market can be summarized as follows. 
\begin{itemize}
\item  Information is incorporated correctly into the price.
\item  Information is incorporated rapidly into the price.
\item  Arbitrage deals profit if there are errors or delays.
\item  Countless traders are always looking for arbitrage opportunities.
\item  The arbitrage opportunity is gradually lost and the market becomes more efficient.
\end{itemize}
Although the standard theory draws a story above, transaction is not established if there is no noise and price will stay distorted if bias is strong.

In an efficient market where noise and bias have no effect and information is perfectly symmetrical, security prices should accurately reflect only information. However, if traders had perfectly and simultaneously symmetrical knowledge and information, including the asset valuation model, the transaction will not function because the traders' valuation of the asset would be the same. They must be doing some rational calculations via intelligence in natural and digital computing. In this sense, a rational representative agent in macroeconomics corresponds to the perfect symmetry of information~(\cite{Lucas1976EconometricPE, Kirman1992WhomOW, 10.2307/2138488}). 

While \cite{black1986noise} treated this ``symmetry breaking'' as noise, the effect of noise on a security's price is expected to be symmetrical based on its nature. But if there exist so many irrational noise traders synchronizing erroneous stochastic beliefs, both affect prices and earn higher expected returns, the unpredictability of noise traders' beliefs creates a risk in the price of the asset that deters rational arbitrageurs from aggressively betting against them. As a result, prices can diverge significantly from fundamental values even in the absence of fundamental risk. Moreover, bearing a disproportionate amount of risk that they themselves create enables noise traders to earn a higher expected return than rational investors do~(\cite{doi:10.1086/261703}).

On the other hand, \cite{doi:10.1126/science.185.4157.1124} found the effect of bias is asymmetric. They described three heuristics that are employed in making judgments under uncertainty: 1) representativeness; 2) availability of instances or scenarios; and 3) adjustment from an anchor. These heuristics are highly economical and usually effective, but they lead to systematic and predictable errors. But the effect of these heuristics has not been detected as a global bias in the security market. Consequently, identifying the specific effects of noise and bias on security price is challenging.

If attention is paid to any statistical property in any complex system, the log-normal distribution is the most natural and appropriate among the standard or ‘‘normal’’ statistics to overview the whole system~(\cite{doi:10.1143/JPSJ.80.072001}). The log-normality emerges as familiar and typical examples of statistical aspects in various complex systems. Since every member of any complex system has its own history, each member is in the process of growth (or retrogression). The log-normal distribution is realized as a result of Gibrat' law, or Mathew effect. It is applied to cities size and growth rate, where proportionate growth process may give rise to a distribution of city sizes that is log-normal. When considering the entire size distribution, not just the largest cities, then the city size distribution is log-normal~(\cite{10.2307/2296055}). However, it has been argued that it is problematic to define cities through their fairly arbitrary legal boundaries. According to \cite{10.1162/003355399556133},
Zipf's law is a very tight constraint on the class of admissible models of local growth. It says that for most countries the size distribution of cities strikingly fits a power law: the number of cities with populations greater than S is proportional to 1/S. Suppose that, at least in the upper tail, all cities follow some proportional growth process (this appears to be verified empirically). This automatically leads their distribution to converge to Zipf's law.

Gibrat's law of proportionate effect also states that the proportional change in the size of a firm is independent of its absolute size. An implication of this is that large and small firms have the same average proportionate rates of growth. Against this law, \cite{10.2307/2296055} shows large firms are growing faster significantly. \cite{10.1257/mac.20150051} construct a tractable neoclassical growth model that generates Pareto's law of income distribution and Zipf's law of the firm size distribution from idiosyncratic, firm-level productivity shocks. Executives and entrepreneurs invest in risk-free assets, as well as their own firms' risky stocks, through which their wealth and income depend on firm-level shocks. By using the model, they evaluate how changes in tax rates can account for the evolution of top incomes in the United States. The model matches the decline in the Pareto exponent of the income distribution and the trend of the top 1 percent income share in recent decades. In the same research direction, \cite{NIREI201625} construct a neoclassical growth model with heterogeneous households that accounts for the Pareto distributions of income and wealth in the upper tail. In an standard Bewley model~(\cite{BEWLEY1977252}), they feature households' business productivity risks and borrowing constraints, which they find generate the Pareto distributions. Households with low productivity rely on wages and returns from safe assets, while high productivity households choose not to diversify their business risks. Their model can quantitatively account for the observed income distribution in the U.S. under reasonable calibrations. Furthermore, they conduct several comparative statics to examine how changes in parameters affect the Pareto distributions. In particular, they find that the change in the top tax rates in the 1980s potentially accounts for much of the observed increase in top income dispersion in the last decades. Their analytical result provides a coherent interpretation for the numerical comparative statics.

In this article, I present the true fundamentals hypothesis based on rational expectations~(\cite{10.2307/1909635}) and, using a log-normal distribution model, detect global bias components from the price-earnings (P/E), price-to-book (P/B), and price-to-cash flow (P/CF) ratios. The traditional theory of the firm is based on the assumption that the firm acts in the stockholders' interests and that stockholders are interested in profit, so that the object of the firm is to maximize profit. However, in fact, there is a certain range in the profit concept~(\cite{10.2307/2977477}). The analysis results strongly support the true fundamentals hypothesis as the detected biases show similar characteristics. Additionally, the results show that the cash flow indicators contain relatively few bias components and are closer to the true fundamentals. I further demonstrate and examine why the positive P/IC ratio among the indicators analyzed is a proxy for the true fundamentals that does not include bias components.

\section{Hypothesis}

When the true fundamentals of listed companies at time $t$ is denoted as $X_t$ and their growth rate is denoted as $R_t$, the growth of those companies can be expressed by the following Gibrat's process
\begin{equation}
 X_t=R_tX_{t-1} \label{1}.
\end{equation}
It is important to note, however, that I assume that the growth rates
$R_t$s are mutually independent random variables that follow the same
distribution with finite variance. The initial value of the fundamentals
set as $X_0$ yields
\begin{equation}
 X_T=X_0\prod_{t=1}^T R_t
\label{2}
\end{equation}
at time $T$. Taking the log of both sides of the equation results in
\begin{equation}
 \log X_T =\log X_0+\log R_1+\cdots +\log R_T.
\label{3}
\end{equation}
Therefore,
\begin{equation}
 \log X_T\sim LN(\mu, \sigma^2)
\label{4}
\end{equation}
would hold true for a sufficiently large $T$ based on the central limit theorem. Essentially, the true fundamentals $X_T$ of listed companies follows the log-normal distribution.

Furthermore, by assuming rational expectations through the future point in time $T$ as of the present point in time 0 on the premise of a going concern, the following equation becomes true:
\begin{equation}
 X_0=\mathbb{E}[X_T]\prod_{t=1}^T \mathbb{E}[R_t^{-1}].
\label{5}
\end{equation}
 Therefore, the rational expectations $X_0$ for true fundamentals also follow the log-normal distribution. In other words, the true fundamentals $X_t$ at a given point in time $t$ follow the log-normal distribution.

\section{Analysis}

\subsection{Data}

Among the actual fundamental indicators, the price-earnings (P/E), forward price-earnings (P/FE), price-to-book (P/B), price-to-cash flows from operating activities (P/OC), price-to-cash flows from investing activities (P/IC), price-to-cash flows from financing activities (P/FC), and the price-to-cash equivalents at the year-end (P/CE) ratios are analyzed. The cash flow ratios were divided into positive and negative data. The daily data with absolute values less than 1,000 were extracted for all companies listed in Japan for the period spanning 1,817 business days from January 2007 to May 2014. 

\subsection{Test}

If the true fundamentals hypothesis is true, the actual fundamental indicators should follow the log-normal distribution. Therefore, the $p$-values were calculated using the Kolmogorov-Smirnov test, Pearson's $\chi^2$ test, and Anderson-Darling test, to test the goodness of fit to the log-normal distribution. It is recognized that the results of these tests are strongly affected by the sample size.

Since the sample size differs by indicator, samples at the 300 quartiles were extracted to calculate the $p$-value for testing the goodness of fit of the overall average of each indicator's samples to the log-normal distribution with variance. In all cases, the null hypothesis is that ``the indicator follows the log-normal distribution,'' or the otherwise worded ``the indicator is a proxy of true fundamentals.'' The indicator reflects the true fundamentals more when the $p$-value is closer to 1. When the $p$-value is closer to 0, the bias is stronger. The lowest $p$-value for each test is shown in Table~\ref{tab1} for the data spanning 1,817 business days.

\begin{figure}[ht]
\begin{center}
\includegraphics[width=11 cm]{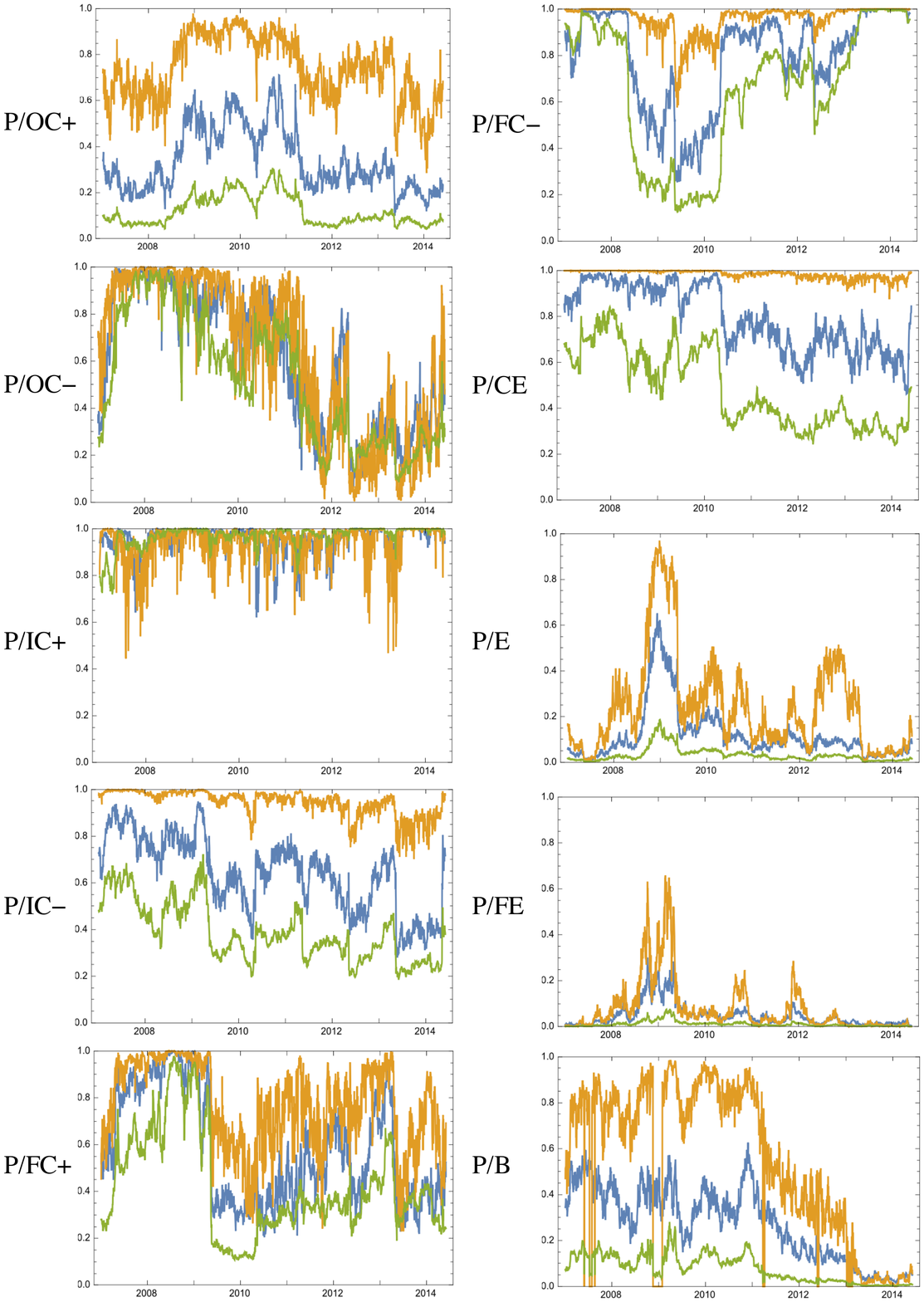}
\caption{Time series of the goodness of fit test to the log-normal distribution. Blue: Kolmogorov-Smirnov, Yellow: Peason's $\chi^2$, Blue: Anderson-Darling. The higher the value, the less bias there is, vice versa. \label{bias1}}
\end{center}
\end{figure}

The test results indicate that bias has a strong effect on the P/E, P/FE, and P/B ratios. It is consistent with our instincts that forward price-earnings ratio (P/FE) is most biased. Interestingly, among all price-to-cash flow ratios, the bias is significant only for P/OC ratio. It can be easier to manipulate than others. I illustrate the time series of $p$-values since the $p$-value is an indicator of the strength of bias. From table~\ref{tab1}, Figure~\ref{bias1} and Figure~\ref{bias2}, it is clear that the positive P/IC ratio is the best proxy for the true fundamentals that does not include bias components. Probably, it should be difficult to manipulate according to the opinions of management and accounting.
\begin{table}[ht] 
\caption{The lowest $p$-values for the data spanning 1,817 business days. Samples at the 300 quantile. Significant (bias) levels: 1\%(***), 3\%(**), 5\%(*)\label{tab1}}
\newcolumntype{C}{>{\centering\arraybackslash}X}
\begin{tabularx}{\textwidth}{llll}
\toprule
\textbf{Index}	&\qquad \textbf{Kolmogorov-Smirnov}	&\qquad\quad \textbf{Peason's $\chi^2$} &\qquad \textbf{Anderson-Darling}\\
\midrule
P/OC$+$ &\qquad\qquad\qquad 0.0986 &\qquad\qquad 0.2843 &\qquad\qquad\qquad 0.0398$^{*}$ \\
P/OC$-$ &\qquad\qquad\qquad 0.0634 &\qquad\qquad 0.0045$^{***}$ &\qquad\qquad\qquad 0.0850 \\
P/IC$+$ &\qquad\qquad\qquad 0.6205 &\qquad\qquad 0.4441 &\qquad\qquad\qquad 0.7201 \\
P/IC$-$ &\qquad\qquad\qquad 0.2787 &\qquad\qquad 0.7012 &\qquad\qquad\qquad 0.1849 \\
P/FC$+$ &\qquad\qquad\qquad 0.2042 &\qquad\qquad 0.2287 &\qquad\qquad\qquad 0.1093 \\
P/FC$-$ &\qquad\qquad\qquad 0.2522 &\qquad\qquad 0.5763 & \qquad\qquad\qquad 0.1269 \\
P/CE &\qquad\qquad\qquad 0.4611 &\qquad\qquad 0.8740 &\qquad\qquad\qquad 0.2376 \\
P/E &\qquad\qquad\qquad 0.0170$^{**}$ &\qquad\qquad 0.0000$^{***}$ &\qquad\qquad\qquad 0.0018$^{***}$ \\
P/FE &\qquad\qquad\qquad 0.0107$^{**}$ &\qquad\qquad 0.0040$^{***}$ &\qquad\qquad\qquad 0.0019$^{***}$ \\
P/B &\qquad\qquad \qquad0.0032$^{***}$ &\qquad\qquad 0.0023$^{***}$ &\qquad\qquad\qquad 0.0005$^{***}$ \\
\bottomrule
\end{tabularx}
\end{table}
\unskip

\unskip

\subsection{Shape of Bias}

\begin{figure}[ht]
\begin{center}
\includegraphics[width=10 cm]{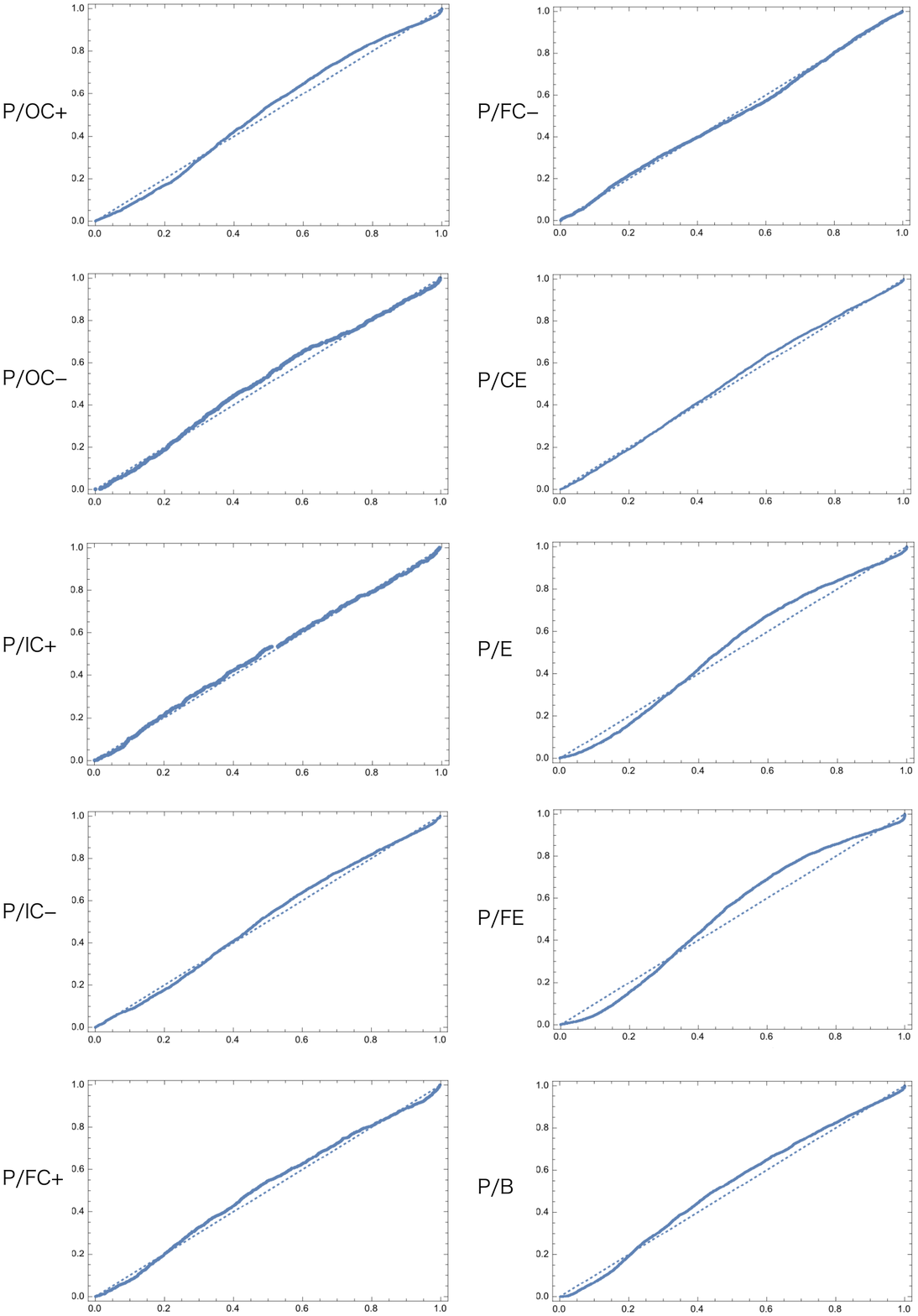}
\caption{The more the data deviates from a straight line showing a log-normal distribution, the stronger the bias.  These graphs are drawn with data from the first day of the time series. The shape of the bias varies proportionally to Figure~\ref{bias1}, but is stable throughout the time. \url{https://figshare.com/projects/Intelligence_and_Global_Bias_in_the_Stock_Market/131711}\label{bias2}}
\end{center}
\end{figure}

The detected biases exhibit similar characteristics. To facilitate an intuitive understanding of these results, each indicator is shown in figure~\ref{bias2} and is compared to the log-normal distribution. While the figures only show the data for the first day of the period, January 4, 2007, the characteristics of the biases themselves remain essentially unchanged throughout the period although the levels fluctuated. It is worth noting that the shape of the curve, and in particular the strength of the bias, is independent of whether the economy is doing well or not. The stock market, which is clearly artificial, appears to have a natural intelligence.

\unskip

\section{Discussion and Conclusion}

The results indicate that the major fundamental indicators including the P/E, P/FE, and P/B ratios are strongly affected by bias. Bias also has a significant effect on the positive and negative P/OC ratios. Additionally, there may be a weak effect of bias on the negative P/IC, positive and negative P/FC, and P/CE ratios.

When the test results are compared, the positive P/IC ratio is the stable proxy of true fundamentals among all these indicators, and it is important to consider why this occurs.

Positive cash flows from investing activities represent the realized gain or loss from past investments such as marketable securities, tangible fixed assets, the sales of investment securities, and income from the collection on loans declared at the end of accounting period. In other words, positive cash flow is the indicator that most directly reflects the past business decisions that determine the growth of a company. The year-end cash equivalent ratio and other cash flow ratios can be interpreted to strongly reflect the true fundamentals because they have less bias components due to the strong characteristic of having definite value.

On the other hand, although the P/E ratio and P/B ratio are definite values, investors might not view them as indicators that reflect the true fundamentals since there is a high degree of freedom in accounting. The P/OC ratio might also have a lower credibility compared to other cash flow indicators.

Where do these biases come from? The answer is extremely simple:  ``Cash is a fact, profit is an opinion.'' Namely, opinions of management and accountant are added as noise to true fundamentals. As a result, Kesten process~(\cite{10.1007/BF02392040})
\begin{equation}
     X_t = R_t X_{t-1}+\epsilon_t,\quad E[\epsilon_t]>0
\end{equation}
is realized and the Pareto distribution is to be obtained. This means their opinions are accompanied by a positive bias.

In fact, these biases fit the Pareto distribution quite well. The generalized Pareto distribution (GPD) is represented by the following functions. 

\paragraph{Generalized Pareto Distribution (GPD)}
\[
F(x)=1-\left(1+\left(\frac{x-\mu}{\kappa}\right)^{1/\gamma}\right)^{-\alpha} 
\]
\[
f(x)=\frac{\alpha}{\gamma}\kappa^{-1/\gamma}(x-\mu)^{-1+\frac{1}{\gamma}}\left(1+\left(\frac{x-\mu}{\kappa}\right)^{1/\gamma}\right)^{-\alpha-1}
\]
\begin{figure}[ht]
\begin{center}
\includegraphics[width=11cm]{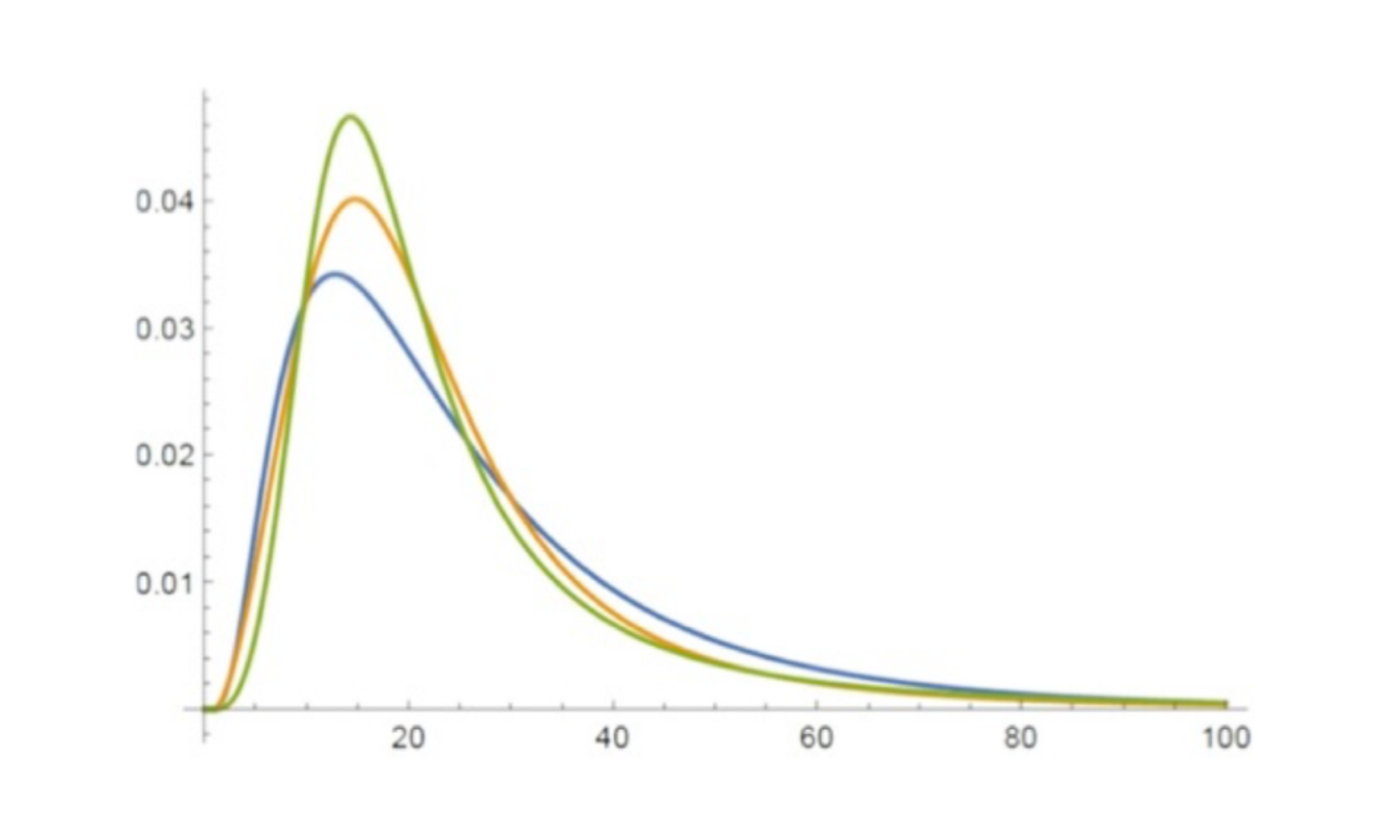}
\caption{Comparison of three distributions: Log-Normal(blue), GDP-1(Yellow), GDP-2(Green)\label{bias3}}
\end{center}
\end{figure}   
\unskip

\begin{figure}[ht]
\begin{center}
\includegraphics[width=13cm]{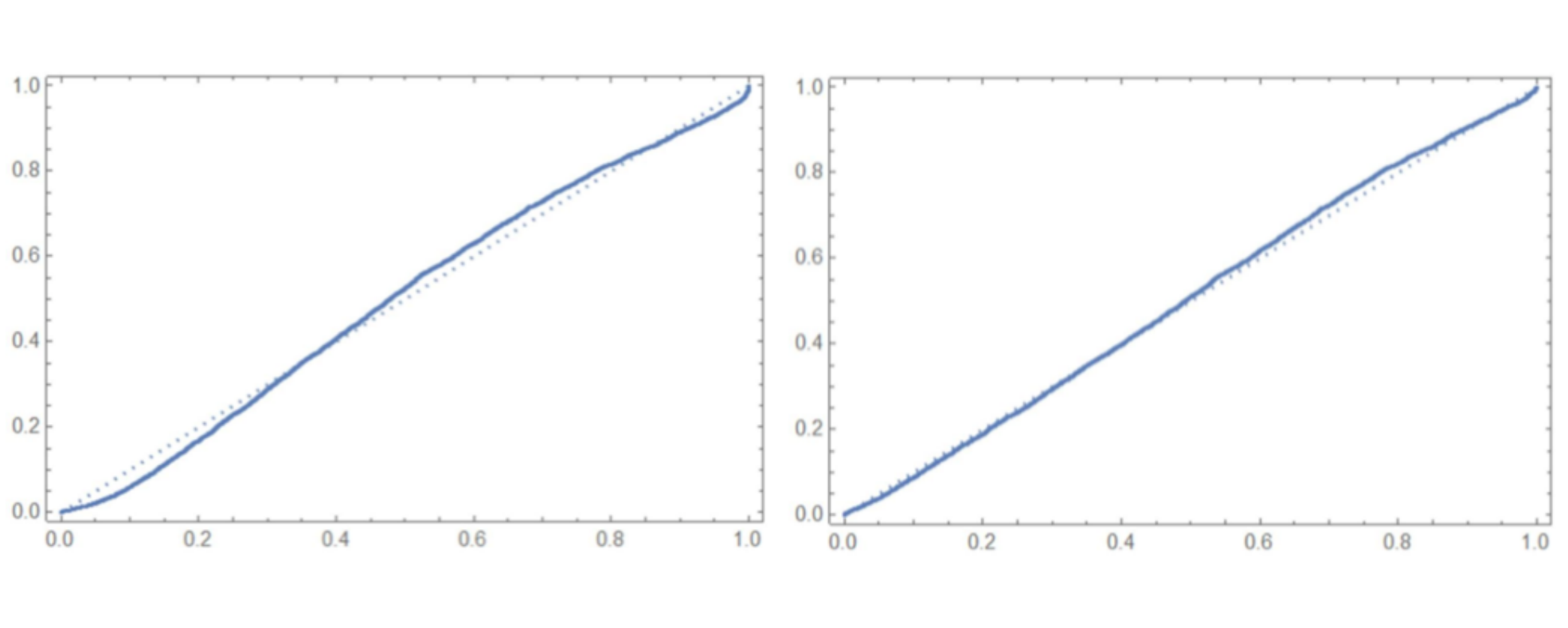}
\caption{Left: GDP-1 to fit P/FE ($\kappa=18.82,\alpha=1, \gamma=0.385, \mu=0.993$),  Right: GDP-2 to fit P/FE ($\kappa=13.70,\alpha=0.515, \gamma=0.238, \mu=0.993$)\label{bias4}. Note that P/FE has the strongest bias.}
\end{center}
\end{figure}

What are the implications of the results? Does the existence of bias negate the efficient market hypothesis? From the view point of intelligence in natural and digital computing, how should we deal with this problem?

The stock market, which is clearly artificial, and where countless traders are using digital computers to trade, appears to have a natural intelligence. Intelligence does not arise only in individual brains; it also arises in groups of individuals~(\cite{malone2015handbook}). If the true fundamentals hypothesis is really true, the major fundamental indicators including the P/E, P/FE, and P/B ratios are strongly affected by bias. This means that investors might not view them as indicators that reflect the true fundamentals since there is a high degree of freedom in accounting. Globally, this implies that the efficient market hypothesis holds~(\cite{fama1965behavior}). The shape of the curve, and in particular the strength of the bias, is stable throughout the time and independent of whether the economy is doing well or not. The curve with bias is the path of the ants~(\cite{kirman1993ants, app12147019}) to avoid obstacles, so to speak, and traders move back and forth along it in response to noise~(\cite{black1986noise, doi:10.1086/261703}). This noise, among other things, plays a role in activating the market system, which knows true fundamentals of firms and represents as a stable global bias in the stock market.

\newpage
\bibliographystyle{unsrtnat}
\bibliography{gb}  






\end{document}